\documentclass{amsart}
\usepackage{graphicx}
\usepackage{epsf}
\def\be{\begin{equation}}
\def\ee{\end{equation}}
\def\ba{\begin{array}}
\def\ea{\end{array}}
\def\bea{\begin{eqnarray}}
\def\eea{\end{eqnarray}}
\def\drm{{\mathrm d}}

\def\rot{{\mathrm{rot}}\,}
\def\div{{\mathrm{div}}\,}

\begin{document}

\vspace{-4truecm} %
{}\hfill{DSF$-$29/2006} %
\vspace{1truecm}

\title[Hole theory and Quantum Electrodynamics in a Majorana's
paper]{Hole theory and Quantum Electrodynamics in an
unknown {\it French} manuscript by Ettore Majorana}%
\author{S. Esposito}%
\address{{\it S. Esposito}: Dipartimento di Scienze Fisiche,
Universit\`a di Napoli ``Federico II'' \& I.N.F.N. Sezione di
Napoli, Complesso Universitario di M. S. Angelo, Via Cinthia,
80126 Napoli ({\rm Salvatore.Esposito@na.infn.it})}%



\begin{abstract}
We give an accurate historical and scientific account of a
previously unknown manuscript written by Ettore Majorana in
French. The retrieved text deals with Quantum Electrodynamics by
using the formalism of field quantization, and it is here
reported, for the first time, in translation. It is likely related
to an invited talk for a conference at Leningrad (or Kharkov) in
1933 (or 1934) which, however, Majorana never attended. Probably
this manuscript refers to the last missing papers of the
``Senatore folder'', given by Majorana to one student of his at
the University of Naples in 1938, just before his disappearance.
\end{abstract}

\maketitle


\section{The last papers written by Majorana}

\noindent Ettore Majorana, born on 5 August 1906, taught from
January to March 1938 at the University of Naples, having obtained
the full professorship in Theoretical Physics ``for high and
well-deserved fame'' \cite{Recami, pw}. The interest in the course
held by the great Italian physicist, the only one effectively
delivered to students \cite{DeGregorio}, has been already pointed
out in the literature (see, for example, Refs. \cite{Moreno},
\cite{Weyl}). It is fundamentally based on the forefront feature
with which Majorana characterized his teaching of Quantum
Mechanics. The historical and personal adventure of Majorana,
instead, is centered around his mysterious disappearance at the
end of March 1938, exactly during his stay in Naples. On Friday 25
March (Majorana held his 21st lesson the day before), according to
one of his student, $\ll$contrary to what he usually did, Majorana
came to the institute and remained there for few minutes... From
the corridor leading to the small room where I was writing, he
called me by name: `Miss Senatore...'. He didn't enter in the room
but remained in the corridor; I went to him, and he gave me a
closed folder, telling me: `that's some papers, some notes, take
them.... We will talk about them later'. Afterwards he went away
and, turning around, said again, `we will talk about them
later'.$\gg$ \cite{Senatore}. That folder contained the lecture
notes for the course on Theoretical Physics prepared by Majorana
who, in that occasion, wanted to give once and for all to one of
his student. What remains of the original papers with those notes
was published some years ago \cite{bibliopolis87}. However, when
Senatore looked at such publication, after very many years since
1938, she pointed out that $\ll$some chapters of the lectures are
missing; their text (completed by the notes of the lecture the
professor would have delivered on the day after his disappearance)
was given to me. A few other sheets are still missing; these were
written in the same original and neat way as the others, but they
do not correspond to the lectures he had already given$\gg$
\cite{Senatore}. The path followed by the original notes by
Majorana from 1938 to 1966, when Edoardo Amaldi deposited them at
the Domus Galilaeana in Pisa together with the other scientific
(unpublished) papers by Majorana, has been quite tortuous, as
pointed out elsewhere \cite{Moreno}. Thus it could be not
surprising that some ``chapters'' have been ``lost''. Nowadays,
however, we may safely assert that the whole set of the lecture
notes for the Theoretical Physics course have been recovered
\cite{bibliopolis06}, thanks to the recent discovery of the Moreno
Paper \cite{Moreno}. This document contains the notes
corresponding to six lectures of the given course, whose original
manuscripts are missing.

Instead, with regard to the notes corresponding to what Senatore
interpreted as ``the lecture the professor would have delivered on
the day after his disappearance'', a very recent analysis
\cite{path} has concluded that they do not refer at all to a
lecture of that course of Theoretical Physics, but rather to the
text of a general conference or seminar probably addressed to the
researchers of the Institute of Physics in Naples itself,
certainly not devoted to students.

\section{The Senatore folder}

\noindent About the remaining ``few other sheets'', strangely up
to now no fruitful research has been conducted, although a long
period elapsed since the time when the Majorana papers were made
publicly available. Such a research is performed now, based mainly
on the papers at the Domus Galilaeana, and the results are here
presented for the first time.

The folder with the Naples lecture notes given to Senatore on 25
March 1938 contains now the following documents.

Firstly, the original of a letter written by Gilberto Bernardini
to Amaldi on 2 December 1064 (reproduced in Ref.
\cite{bibliopolis87}) is present in it. In this letter Bernardini
replied to Amaldi about his persistent request on the Majorana
lecture notes. Evidently, such a letter was added by Amaldi in
recent times (around 1966) to the folder.

The second document\footnote{The subdivision in terms of
``documents'' is made here for convenience, but it does not
correspond to any real division. Indeed, the folder contains only
spare sheets, according to what here described.} present in the
folder is composed by 22 A4 papers (each of one corresponding to 4
pages). They are the original notes corresponding to 10 lectures
of the Majorana course \cite{bibliopolis06}.

The third document is, instead, composed of only 3 papers (each of
one corresponding again to 4 pages). The text contained in it
refers to what mentioned above regarding the general conference
delivered by Majorana (see Ref. \cite{path}). For possible future
reference, this document of 3 papers will be denoted as  ``{\it
excerpta} Senatore I''.

Unexpectedly, the folder at the Domus Galilaeana contains also
another document, composed of 2 A4 papers (the first one made of 4
pages, while the other one of only two pages). Such a document,
denoted here as ``{\it excerpta} Senatore II'', is completely
different from the {\it excerpta} Senatore I (and from the lecture
notes), due to peculiarity of the text, which is written in {\it
French} rather than in Italian, and to particular writing details.

The content of the text of this manuscript, reported (translated
into English) in the appendix, concerns about an application of
the formalism of second quantization to the Dirac's hole theory
({\it excerpta} Senatore IIa) and some topics of Quantum
Electrodynamics ({\it excerpta} Senatore IIb e IIc), which will be
discussed from a theoretical point of view in the following
section. However, it is interesting to observe at once that,
despite the apparent variety of topics, the whole French text
present in the 2 papers seems to have been written in no more than
one occasion.

\section{The reference theoretical background}

\noindent In the general framework of Wave Mechanics, basically
introduced by Erwin Schr\"o\-din\-ger, the existence of a ``matter
quantum'', such as the electron, posed the same theoretical
problems of those implied by the existence of light quanta in the
context of the Maxwell electromagnetic theory. This was clearly
noted by Werner Heisenberg in 1929 \cite{Hei29}. Indeed, if the
fundamental problem related to the existence of the photons was
that of explaining the interference phenomenon, for the electrons
it was that of deducing the quantization of the electric charge
starting from the Schr\"odinger wavefunction.

As well known, the solution to such difficulties came out with the
introduction of the field quantization (or ``second
quantization'') procedure that, after a preliminary work by
Pascual Jordan in 1926 \cite{Jordan26}, it was applied for the
first time by Paul A. M. Dirac in 1927 \cite{Dirac27} to the
quantization of the electromagnetic field. The following step was
that of considering particles with non vanishing mass, and this
was achieved by Jordan and Oskar Klein \cite{JK27}, who
generalized the Dirac quantization method to bosonic fields. The
procedure to be adopted for fermions, according to the Pauli
exclusion principle, required instead the ``invention'' of
anticommutators, as pointed out by Jordan and Eugene Wigner in
1928 \cite{JW28}.

The key point was, evidently, the realization that the
(Schr\"odinger, Dirac or Maxwell) wave field does not represent a
single particle but, on the contrary, an arbitrary number of
particles. To this regard, it is particularly expressive the title
given by Jordan and Klein to their fundamental paper, {\it On the
many-body problem of the Quantum Theory}.

The inclusion of the interaction between light and matter in the
second quantization procedure was systematically considered in
1929-1930 by Heisenberg and Wolfgang Pauli \cite{HP29}, who
elaborated a first theory of Quantum Electrodynamics by using the
lagrangian formalism. In this theory, however, some fundamental
problems began to impose to the attention of the theorists, such
as the emergence of an infinite Coulomb self-energy for the
electron and the problem of the negative-energy states (that is
the transition between states with energy $E=+ m c^2$ to states
with $E=- m c^2$). Despite the engagement of some of the best
theoretical physicists of that epoch, such problems remained
unsolved for a given period of time. In particular the problem of
the negative energy states increased when Ivar Waller
\cite{Waller30} proved that they were effectively required in
order to obtain the Thomson classical limit of the scattering
cross section of X-rays from atoms. The Dirac theory of ``holes''
\cite{Dirac30}, interpreted earlier in terms of protons (by
following an Hermann Weyl's conjecture \cite{Weyl29}) and then
assuming the existence of positrons as subsequently observed,
offered a simple and interesting solution to the problem, as well
known. It was thus introduced the concept of ``vacuum'' (or, using
the words by Dirac, ``the normal state of electrification'') as a
sea of electrons where all the positive energy states are empty,
whilst those with negative energy are all occupied.

The general acceptance of the hole theory was, initially, not
unquestioned inside the physicists' community, and still in 1933,
after the experimental discovery of the ``antielectron'', Pauli
expressed in many occasions his perplexity  about it
\cite{Pauli33}. A completely different viewpoint was instead
adopted by Heisenberg who was favorably impressed by the
substantial symmetry, in the Dirac theory, between processes
involving electrons and those involving positrons \cite{HS33}, an
idea that will be later formally recovered  by Majorana in his
famous article on the {\it Symmetrical theory of electrons and
positrons} \cite{elpos} (whose genesis could, probably, be dated
to 1933, when Majorana visited Heisenberg in Leipzig). Heisenberg
himself, in fact, in 1931 elaborated a particularly interesting
(for our aims) application, in which he considered the symmetry
between holes and electrons in an occupied atomic level or in an
occupied energy band of a crystal \cite{Hei31} (that is: an
application to an actual case, in contrast to the ``vacuum'' by
Dirac). In this paper Heisenberg obtained a second quantized
hamiltonian expressed in terms of holes (see Eq. (15) in Ref.
\cite{Hei31}), where some additional one-particle operator terms
comes out from the commutation between two-particle operators. The
application to the case of an atomic level of $N$ electrons
containing, however, only $n$ electrons was studied in the first
part of the article, and led to a wave equation (Eq. (17) in Ref.
\cite{Hei31}) for $N-n$ holes. In the second part of it, the
electron-hole symmetry in the case of a metal with an
``anomalous'' Hall effect is instead considered. As already
mentioned, Heisenberg used the formalism of second quantization,
by making explicit reference to both the general article by Jordan
and Klein of 1927 \cite{JK27} and to that of Jordan and Wigner of
1928 \cite{JW28}, where the anticommutators for fermions (namely,
electrons and holes) were properly introduced.

In the {\it excerpta} Senatore IIa Majorana substantially followed
the track left by Heisenberg in \cite{Hei31}, but it is not clear
if the author referred to an actual case (as that of Heisenberg)
or, even more interestingly, if his intention was that of
formulating a general theory of holes. Indeed, in the opening of
the manuscript, the general assumptions to which Majorana appealed
were very clearly and explicitly asserted, together with the
reference point of the method of the ``quantization of the
Schr\"odinger equation'' (i.e. that of second quantization). As
well as Heisenberg, Majorana obtained the expression of the
hamiltonian of the system in terms of holes, but it is evident
that the theory elaborated in this manuscript is incomplete (or,
rather, interrupted)

It is also particularly interesting the ``justification'' that
Majorana gave for the adoption of anticommutators instead of
commutators, which he referred to the particular form of the
hamiltonian and to the corresponding equations of motion to be
satisfied.

In the {\it excerpta} Senatore IIb and IIc,\footnote{Although the
same topic is treated in these two pages, it is preferable to
introduce a further subdivision into two parts, since it is
evident the lack of connection between them.} instead, Majorana
considered a topic that he frequently studied in his personal
research notebooks \cite{Quaderni}: if, as recalled above, the
Maxwell theory of electromagnetism should be viewed as the Wave
Mechanics of the photon, then it should be possible to write the
Maxwell equations as a Dirac-like equation for a suitable
wavefunction. An alternative model to the theory of
Electrodynamics, based on an analogy with the Dirac theory of the
electron, was already proposed by Oppenheimer in 1931
\cite{Opp31}; the contribution given by Majorana to this subject
has been already discussed elsewhere \cite{RecEspGian}, and we
refer there the interested reader for further inputs related to
the second part of the manuscript considered.

Regarding the {\it excerpta} Senatore IIc it is instead
particularly interesting to point out the clear and explicit
presentation of the experimental properties of photons, upon which
the novel formulation of Electrodynamics may be based, related to
the velocity, energy, momentum and spin of photons. Note also that
the Dirac-like equation for the photons reported at the end of the
IIc manuscript is substantially {\it different} from that
considered in the IIb one (and by Oppenheimer). It is applied to a
{\it two-component} wavefunction (corresponding to the only two
different polarization states of the photon) instead of a
three-component one (corresponding to the electric and magnetic
fields of the associated wave).

\section{The composition of the manuscript and a Soviet conference}

\noindent Unfortunately, the {\it excerpta} Senatore are not
dated, so that it is difficult, without suitable assumptions, an
accurate analysis of the genesis and composition of the Majorana
work here considered. However, some ``reasonable'' considerations
may be usefully developed, without forgetting the fundamental
point just pointed out.

First of all, based on the kind of writing used, we can certainly
exclude a composition date around that for the university lectures
at Naples (1938). Indeed, from a comparison with other documents
present in the archive of the Domus Galilaeana, especially the
original manuscripts of the published articles by Majorana, we
deduce that some writing details are very similar to those found
for the paper in Ref. \cite{atomi} of 1932.\footnote{They are also
similar to those of the article in Ref. \cite{sdoppiamento} of
1928, which would denote a composition preceding the Heisenberg
paper of 1931 \cite{Hei31}. Although such an hypothesis cannot be
completely excluded, based on what noted in the previous section
and to what discussed below, we consider it not very likely.}

The analysis of the specific contents of the Majorana manuscript
seems to confirm a composition date not prior than 1932 revealing,
for the {\it excerpta} Senatore IIa, a certain dependence on the
Heisenberg article in \cite{Hei31} and, for the {\it excerpta}
Senatore IIb e IIc, at least the knowledge of the Oppenheimer
paper of the same year \cite{Opp31}.

These only data, however, lead to generic conclusions, pointing
out only a lower limit on the composition date, and certainly they
do not offer hints on the genesis and the expected use of the
manuscript.

Looking at the whole set of papers left by the Italian physicist,
conserved not only in public archives like the Domus Galilaeana,
but also in the private one of the Majorana family, we have found
that {\it only} another document was written in French. It
corresponds to a draft of a letter by Majorana in replying to an
invitation to a conference (see the document MX/R1 in
\cite{Recami}). In the following we give the translation of it:
\begin{quote}
Dear Sir,

I thank you warmly for your invitation to participate to the next
congress that will take place in Leningrad. I am happy to accept
and to have the occasion of visiting, at the same time, your big
and beautiful country. I have talked about your invitation with
Mr. Fermi and Mr. Rossi. Fermi is engaged in a course of
conferences in America and cannot participate. Rossi, on the
contrary, assured me that will accept with pleasure to go to
Russia.

Dear Sir, receive my hearty greetings together with my warmest
thanks.

Yours

{${}$}\hfill{Ettore Majorana}

${}$

{\begin{tabular}{l} Ettore Majorana \\
Institute of Physics of the R. University \\
Rome \\ ${}$ \\
Bruno Rossi \\
Institute of Physics of the R. University \\
Padua \end{tabular}}\hfill{}
\end{quote}
Although even this document is not dated, we can deduce some
interesting information from what reported in it, if properly
examined.

First of all, the letter containing the invitation to the
conference was directed to Majorana as well as to Bruno Rossi (and
Enrico Fermi), who worked at that time at the Institute of Physics
in Padua. Now, it is known that Rossi was in Padua since the Fall
of 1932, and that in between 1932 and 1933 he worked together with
Fermi (in Rome) on some questions about Cosmic Rays Physics (see
the paper in \cite{RossiFermi}  and its presentation in the book
\cite{FNM} at page 509).

The second important fact regards the mentioned conference,
apparently to be held in Leningrad. In the years considered, the
only one conference of interest for us is the First All-Union
Conference on Nuclear Physics \cite{allunion}, whose main topic
was the Physics of atomic nucleus. It was held on 24-30 September
of 1933 at the Leningrad Physico-Technical Institute
\cite{Vizgin}. Contrary to what asserted in the letter above,
neither Majorana nor Rossi participated to such a conference (as
well as to other ones in the Soviet Union) while, among the others
(for example, I.E. Tamm, V.A. Fock, G.A. Gamov, P.A.M. Dirac, F.
Joliot, V.F. Weisskopf) Franco Rasetti, another member of the
``Via Panisperna boys'' group, took part to it \cite{Vizgin}. It
is even interesting to observe that Dirac participated to this
conference with a talk on the ``Theory of the positron''
\cite{positrona}, that is exactly the hole theory considered by
Majorana in the mentioned paper.

The date of the Leningrad conference is in agreement also with
another important piece of information contained in the Majorana
letter, regarding the absence of Fermi who was ``engaged in a
course of conferences in America''. Indeed, in the August 1933
Fermi, accompanied by Emilio Segr\`e, went to the summer school at
Ann Arbor \cite{Segre} \cite{FNM77b}, where he was invited to
deliver some lectures (the same happened in 1930 and 1935).

Nevertheless the hypothesis just considered is not free from some
difficulties, concerning mainly {\it who} effectively suggested to
invite Majorana and Rossi to the conference and {\it when} the
replying letter was effectively written by Majorana, after that
the invitation arrived in Rome. In fact, regarding the first
point, the work by Majorana (and, partially, that of Rossi) was
not yet sufficiently known by the international community at the
beginning of 1933. Instead, for the second point, we have to
recall that Majorana was in Leipzig (and in Copenhagen, but not in
Rome) from January to the beginning of August of 1933
\cite{Recami}, except for the Easter holidays ranging from the end
of April to the beginning of May of that year. Then it seems
unlikely that a visit to Leningrad could be organized only one
month before the conference (in August) or in the short period of
the Easter holidays.

The international fame of Majorana in the field of Nuclear Physics
and, more in general, in that of Theoretical Physics, increased
only after his stay in Leipzig. Incidentally, we note that
Heisenberg publicized a lot the Majorana work (especially that on
the nuclear exchange forces \cite{kern}) in the conferences
following the mentioned visit \cite{degneutro} and, in particular,
in the Solvay Congress at Bruxelles of the October 1933, where
talks were given in French. The invitation to Majorana and Rossi
(Heisenberg himself, who worked in the field of Cosmic Rays
Physics too, was as well a quite good admirer of the Rossi's
papers on this) could then be conceived in this occasion, and then
formalized later.

Taking for grant such an hypothetical occurrence, the conference
cited in the Majorana letter cannot be that of September 1933 in
Leningrad. The subsequent conference in soviet soil was not given
in this town, but in the other important center of Kharkov in May
1934 \cite{Tamm} \cite{cold}. It was, nevertheless, organized by
important Russian physicists working in Leningrad, such as A.F.
Joffe and others. Few non-Russian physicists took part to this
International Conference of Theoretical Physics (then, a
conference not explicitly limited to Nuclear Physics); among the
others, we mention Niels Bohr, Leon Rosenfeld (both were known to
Majorana since its 1933 visit to Copenhagen), Ivar Waller and
Walter Gordon. It is also interesting to note that in the year
preceding this conference, the physicist Victor F. Weisskopf
worked in the Institute of Kharkov \cite{Weisskopf} with Lev D.
Landau and others, while Rudolph Peierls usually visited Landau in
Leningrad \cite{Peierls}. Both Weisskopf and Peierls knew quite
well Majorana \cite{Recami}, the first one was met in Leipzig in
1933 (in this occasion the pair discussed on Quantum
Electrodynamics), while the second one was met in Rome between the
end of 1932 and the beginning of 1933, just before the visit of
Majorana in Germany.

Note that even the absence of Fermi mentioned in the Majorana
letter may be easily accounted by the hypothesis that the
considered conference was that of Kharkov. Indeed, it is known
that in the Summer of 1934 Fermi went to South America, were he
held a series of conferences in Buenos Aires, Montevideo and
elsewhere \cite{Segre}. If we let to interpret literally the words
used by Majorana, the ``course of conferences'' seems to be more
easily referred to that of 1934 in South America than to that of
1933 in the United States.

Summing up, the likely hypothesis on the date of writing of the
mentioned Majorana letter in French are only two: the first one
refers to the Nuclear Physics conference at Leningrad in September
1933, while the second one refers to that of Theoretical Physics
at Kharkov in May 1934.

In both cases, the open problem is the possible relation between
such an invitation letter to a conference in the Soviet Union and
the writing of the manuscripts in the {\it excerpta} Senatore II.
Unfortunately, to this regard, no conclusive supporting evidence
exists. However, by analyzing carefully the life of Majorana
\cite{Recami}, no other occasion seems to imply the possible usage
of the French as possible communication medium, so that it would
be likely that some a link between the two documents exists. If
this conjecture is confirmed, then we can reasonably conclude that
the text present in the {\it excerpta} Senatore II should be
prepared for a possible Majorana talk at the conference in
Leningrad or in Kharkov, although he never took part to it. And,
probably, his non-participation may be well related to the
uncompleted development of the theory present in the {\it
excerpta} that, as mentioned above, is only sketched.

\section{Conclusions}

\noindent From a recent recognition at the Domus Galilaeana in
Pisa we have realized that in the folder of the Majorana lecture
notes, given by Majorana to one of his student in Naples in the
March 1938, just before his disappearance, some spare papers are
present in it, not belonging to the collection of the given notes.
The existence of such papers was mentioned directly by that
student, Gilda Senatore, some years ago, but till now their
effective presence has not been effectively reported. In the
present work we have thus performed, for the first time, an
accurate analysis of the cited manuscript ({\it excerpta} Senatore
II), written by Majorana in {\it French}. It is the only
scientific manuscript written by the author in that language. The
content of it sketches a theory about Quantum Electrodynamics,
with the use of the field quantization formalism. In particular,
he considers some questions regarding the hole theory and the
formulation of Electrodynamics suggested by Oppenheimer in analogy
to the relativistic Dirac theory of the electron.

From what discussed above, it seems likely to suppose that the
text considered here was elaborated in 1933-1934, probably for a
conference in the Soviet Union (at Leningrad in 1933 or Kharkov in
1934) to which Majorana, though invited to participate, did not
take part to it. In any case, we can reasonably exclude a date of
composition far from the period when Majorana stayed in Leipzig
with Heisenberg, since it seems well documented a certain
dependence of the Majorana paper on an article by the German
physicist.

On the other hand, as noted by Weisskopf, who knew Majorana in
Leipzig, $\ll$at that time I was still interested in Quantum
Electrodynamics, and there were two problems to be faced: one was
the problem of the positron, whether it really is contained within
the Dirac equation, the problem of the charge conjugation
symmetry, as we say today, and two, the problem of the nuclear
force, the beginning of Nuclear Physics... All discussions focused
around nuclear structure on the one hand and Quantum
Electrodynamics on the other$\gg$ \cite{Weisskopf}.

Then Majorana, as well as other brilliant theoreticians of that
period, studied actively both fundamental questions \cite{kern}
\cite{elpos}, and the {\it excerpta} Senatore II are a further
interesting proof. Nevertheless the fact remains that the theory
developed in these excerpta is only sketched and, at the moment,
it is not safely to conjecture on why the great theoretical
physicist has not completed his work. Certainly, it is intriguing
that probably Majorana kept with him such papers in Naples after
several years since their writing, and that he likely gave to his
student Senatore before his mysterious disappearance. Further
future researches in this directions may probably put new light on
this novel and fascinating question.

\section*{Acknowledgments}

\noindent The constant interest and the fruitful help given to me
by E. Recami and A. De Gregorio deserve my sincere gratitudr,
which is here fully expressed.

\newpage

\appendix

\section{The text by Majorana}

\noindent In the following we report the English version of the
text by Majorana identified as {\it excerpta} Senatore II. We
point out that, apparently, the author has probably written the
original text in Italian and then translated it into French or,
alternatively, he did not elaborate directly the text by following
the composition rules of the French language. In few places, when
strictly required by clarity, we have added additional explanatory
text between brackets [...].

\bigskip \bigskip

\centerline{\sc {\it Excerpta} Senatore IIa}

\bigskip

\noindent Let us consider a system of $p$ electrons and put the
following assumptions: 1) the interaction between the particles is
sufficiently small allowing to speak about individual quantum
states, so that we may consider that the quantum numbers defining
the configuration of the system are good quantum numbers; 2) any
electron has a number $n>p$ of inner [energetic] levels, while any
other level has a much greater energy. We deduce that the states
of the system as a whole may be divided into two classes. The
first one is composed of those configurations for which all the
electrons belong to one of the inner states. Instead the second
one is formed by those configurations in which at least
\underline{one} electron belongs to a higher level not included in
the $\underline{n}$ levels already mentioned. We will also assume
that it is possible, with a sufficiently degree of approximation,
to neglect the interaction between the states of the two classes.
In other words we will neglect the matrix elements of the energy
corresponding to the coupling of different classes, so that we may
consider the motion of the $p$ particles in the $n$ inner states,
as if only these states exist. Then, our aim is to translate this
problem into that of the motion of $n-p$ particles in the same
states, such new particles representing the holes, according to
the Pauli principle.

In order to reach our purpose, it is worth to adopt the formalism
used in the method known as that of the quantization of the
Schr\"odinger equation. Let us consider the Schr\"odinger
equation:
\[
\dot{\psi} = - \frac{2 \pi i}{h} \, H \psi
\]
as defining a classical field, taking into account the fact that
$\psi$ does not represent a single particle but, rather, a very
large number of particles, in order to neglect the granular
structure of matter. Then, it is useful to highlight such a part
depending on the mutual interaction of the electrons in the
potential appearing in the hamiltonian.
Let us then put: %
\be \label{1} %
- \frac{h}{2 \pi i} \dot{\psi} = H \psi + V \psi ,
\ee %
being \setcounter{equation}{0}
\renewcommand{\theequation}{\arabic{equation}$'$}
\be \label{1p}%
V(P) = \int G(P,P') \,\, \psi^\ast(P') \psi(P') \, \drm \tau ,
\ee %
\setcounter{equation}{1}\renewcommand{\theequation}{\arabic{equation}}where
$G(P',P)$ is the potential of the forces acting between two
particles located in $P$ and $P'$. In a natural way we will
consider the energy of the field as given by the expression%
\bea %
& & \int \psi^\ast \, H \, \psi \, \drm \tau + \frac{1}{2} \int
\psi^\ast \, V \, \psi \, \drm \tau \nonumber \\
& & = \int \psi^\ast \, H \, \psi \, \drm \tau + \frac{1}{2} \int
\!\!\!\! \int G(P,P') \, \psi^\ast(P) \, \psi^\ast(P') \, \psi(P)
\, \psi(P') \, \drm \tau \, \drm \tau' . \label{2}
\eea %
Let us now expand by using a set of orthogonal
functions:\footnote{In the original manuscript, the equations
appearing in the following second and third line are not written
just after the first equation, but in the top corner of the sheet.
In the second equation the symbol $\psi_k$ is used in place of
$\varphi_k$.}
\[
\psi = \sum a_i \, \varphi_i , \qquad \qquad \qquad \int
\varphi_i^\ast \, \varphi_k \, \drm \tau = \delta_{ik} ,
\]
\[
H \varphi_k = \sum_i H_{ik} \, \varphi_i ,
\]
\[
V_{ik} = \frac{1}{2} \sum_{\ell , m} O_{i \ell, k m} \,
a^\ast_\ell a_m ;
\]
we can write:\footnote{The following equation is written, in the
original manuscript, as \[ H_{ik} = \int \widetilde{\psi}_i \, H
\, \psi_k \, \drm \tau . \]}
\[
H_{ik} = \int \varphi^\ast_i \, H \, \varphi_k \, \drm \tau ,
\]
\[
O_{i \ell , k m} = \int \!\!\!\! \int G(P,P') \, \varphi^\ast_i(P)
\, \varphi^\ast_\ell(P') \, \varphi_k(P) \, \varphi_m(P') \, \drm
\tau \, \drm \tau' .
\]
Then by substituting these into the equations of motion we  write:%
\be \label{3} %
\dot{a}_i = - \frac{2 \pi i}{h} \left\{ \sum_k H_{ik} \, a_k +
\sum_{\ell, k, m} O_{i \ell, k m} \, a^\ast_\ell a_k a_m \right\}
,
\ee %
and, as the expression of the energy: %
\be \label{4} %
W = \sum_{i, k} H_{ik} \, a^\ast_i a_k + \frac{1}{2} \sum_{i,
\ell, k, m} O_{i \ell, k m} \, a^\ast_i a^\ast_\ell a_k a_m .
\ee %
Taking into account these equations, we will now give them a
quantum meaning by setting %
\be \label{5} %
\dot{a}_i = - \frac{2 \pi i}{h} \left( a W - W a \right) , \qquad
\qquad \dot{a}^\ast_i = - \frac{2 \pi i}{h} \left( a^\ast W - W
a^\ast \right) ,
\ee %
the quantities $a$ being now matrices. In order that the equations
(\ref{5}) be equivalent to equations (\ref{3}), we easily see that
the quantities $a$ should satisfy the exchange relations:
\begin{eqnarray*}
& & a_i a^\ast_k - a^\ast_k a_i = \delta_{i k} ,\\
& & a_i a_k - a_k a_i = 0 , \\
& & a^\ast_i a^\ast_k - a^\ast_k a^\ast_i = 0 .
\end{eqnarray*}
This means to quantize [the theory] according to the classical
Heisenberg rules since, indeed, the momenta conjugate to the
variables $a$ are classically the quantities $a^\ast$ multiplied
by the factor $-h / 2 \pi i$. The Heisenberg exchange relations
will lead us to consider particles obeying to the Bose statistics,
while we are interested in the other case, namely that of
particles obeying the Fermi statistics. As proved by Jordan and
Wigner, to this end we have to change the signs in the Heisenberg
relations:
\begin{eqnarray}
& & a_i a^\ast_k + a^\ast_k a_i = \delta_{i k} , \nonumber \\
& & a_i a_k + a_k a_i = 0 , \label{6} \\
& & a^\ast_i a^\ast_k + a^\ast_k a^\ast_i = 0 . \nonumber
\end{eqnarray}
This cannot be justified on general grounds, but only by the
particular form of the hamiltonian. In fact, we may verify that
the equations of motion are satisfied to the best by these last
exchange relations rather than by the Heisenberg ones. We now
consider a suitable solution.%
\[
I_k = \left| \ba{cc} 1 & 0 \\ 0 & 1 \ea \right| , \qquad  I'_k =
\left| \ba{cc} 1 & 0 \\ 0 & -1 \ea \right| ; \qquad \alpha_k =
\left| \ba{cc} 0 & 1 \\ 0 & 0 \ea \right| , \qquad \alpha^\ast_k =
\left| \ba{cc} 0 & 0 \\ 1 & 0 \ea \right| ;
\]

\[
a_i = I'_1 \times I'_2 \dots I'_{i-1} \times \alpha_i \times
I_{i+1} \times I_{i+2} \dots ,
\]

\[
a^\ast_i = I'_1 \times I'_2 \dots I'_{i-1} \times \alpha^\ast_i
\times I_{i+1} \times I_{i+2} \dots ;
\]

\[
\alpha^\ast_i \alpha_i = \left| \ba{cc} 0 & 0 \\ 0 & 1 \ea \right|
,
\]
[whose eigenvalues are] 0,1.

\[
\int \psi^\ast \psi \, \drm \tau = \sum_i \alpha^\ast_i \alpha_i =
n .
\]

\noindent \footnote{In the remaining part of the manuscript, some
attempts are carried out by the author in order to calculate the
matrix elements of the energy $W$ (and, in particular, of $H_{ik}$
and $O_{i \ell, k m}$) in the basis of the occupation numbers
$n_1, n_2, \dots, n_i, \dots, n_k, \dots$. Here we do not report
such attempts, which are difficult to interpret.}

\noindent [By setting] $b_i=a_i^\ast$, $b_i^\ast=a_i$, [the
relation]

\[
a_i^\ast a_i + a_i a_i^\ast = 1
\]

\noindent [can be written as]

\[
a_i^\ast a_i + b_i^\ast b_i = 1 ,
\]

\noindent [that is]

\[
a_i^\ast a_i = 1 , \qquad \qquad b_i^\ast b_i = 0 , 
\]

\noindent [or]

\[
a_i^\ast a_i = 0 , \qquad \qquad b_i^\ast b_i = 1 .
\]

\noindent [For the energy $W$ we then have:]

\[
\sum H_{ik} \, a^\ast_i a_k + \frac{1}{2} \sum O_{i \ell, k m} \,
a^\ast_i a^\ast_\ell a_k a_m ,
\]

\noindent [or]

\[
\sum H_{ik} \, b_i b^\ast_k + \frac{1}{2} \sum O_{i \ell, k m} \,
b_i b_\ell b^\ast_k b^\ast_m .
\]

\noindent [From the relation]

\begin{eqnarray*}
b_i^\ast b_k + b_k^\ast b_i = \delta_{ik} , \\
b_i^\ast b_k = \delta_{ik} - b_k^\ast b_i ,
\end{eqnarray*}

\noindent [the first term of the energy is written as]

\[
\sum H_{ii} - \sum \overline{H}_{ik} b_i^\ast b_k .
\]

\noindent [For the second term, by using:]

\begin{eqnarray*}
b_i b_\ell b_k^\ast b_m^\ast &=& b_i \left( \delta{\ell k}-
b_k^\ast b_\ell \right) b_m^\ast \\
&=& \delta_{\ell k} b_i b_m^\ast - b_i b_k^\ast b_\ell b_m^\ast = \dots \\
&=& b_k^\ast b_m^\ast b_i b_\ell + \delta_{\ell k} b_i b_m^\ast +
\delta_{i m} b_\ell b^\ast_k - \delta_{i k} b_\ell b_m^\ast -
\delta_{\ell m} b_i b_k^\ast ,
\end{eqnarray*}

\noindent \footnote{In the last relation the author has omitted
the term $\delta_{\ell m} \delta_{ik} - \delta_{i m} \delta_{\ell
k}$ which, in the sum appearing in the expression of the energy,
leads to a vanishing contribution.}

\noindent [we get:]

\begin{eqnarray*}
\frac{1}{2} \sum O_{i \ell,km} \, b_i b_\ell b_k^\ast b_m^\ast &=&
\frac{1}{2} \sum O_{i \ell,km} \, b_k^\ast b_m^\ast b_i b_\ell \\
&+& \frac{1}{2} \sum O_{i \ell,\ell m} \, b_i b_m^\ast +
\frac{1}{2} \sum O_{i \ell,k i} \, b_\ell b_k^\ast + \dots .
\end{eqnarray*}

\noindent \footnote{The text contained in the original manuscript
ends with few other calculations which we do not report.}

\bigskip \bigskip

\centerline{\sc {\it Excerpta} Senatore IIb}

\bigskip

\noindent These\footnote{The beginning of this part of the
manuscript is evidently the continuation of a previous one, which
is lost. However, see what the author writes in the following}
equations, however, satisfy a fourth one. \\
Then Opp.\footnote{Probably, Majorana refers here to the paper by
R.J. Oppenheimer in {\it Phys. Rev.} {\bf 38} (1931) 725.} has
attempted to build a theory with three vector components.

\noindent [The Maxwell equations in vacuum:]

\[
\frac{1}{c} \frac{\partial E}{\partial t} = \rot H , \qquad
\frac{1}{c} \frac{\partial H}{\partial t} = - \rot E ,
\]
\[
\div E = 0 , \qquad \div H = 0 ,
\]

${}$

\noindent [by putting]

\begin{eqnarray*}
\psi_x &=& E_x - i H_x , \\
\psi_y &=& E_y - i H_y , \\
\psi_z &=& E_z - i H_z ,
\end{eqnarray*}

${}$

\noindent [may be written in the following form:]

\[
\frac{1}{c} \frac{\partial \psi}{\partial t} = \rot \left( H + i E
\right) = i \, \rot \left( E - i H \right) = i \, \rot \psi ,
\]

\[
\left\{ \ba{l} \displaystyle \frac{1}{c} \frac{\partial
\psi}{\partial t} - i \, \rot \psi = 0 , \\ \\ \displaystyle \div
\psi = 0 , \ea \right.
\]

${}$

\[
\left\{ \ba{l} \displaystyle \frac{1}{c} \frac{\partial
\psi_x}{\partial t} - i \, \frac{\partial \psi_z}{\partial y} + i
\, \frac{\partial \psi_y}{\partial z} = 0 ,
\\ \\
\dots \\ \\
\displaystyle \frac{\partial \psi_x}{\partial x} + \frac{\partial
\psi_y}{\partial y} + \frac{\partial \psi_z}{\partial z} = 0 . \ea
\right.
\]

${}$

\noindent [By introducing the operators]

\[ W = - \frac{h}{2 \pi i} \, \frac{\partial ~}{\partial t}, \qquad
p_x = \frac{h}{2 \pi i} \, \frac{\partial ~}{\partial x}, \qquad
\dots \]

${}$

\noindent [we have:]

\[
\left\{ \ba{l} \displaystyle \frac{1}{c} W \psi_x + i \, p_y
\psi_z - i \, p_z \psi_y = 0,
\\ \\
\displaystyle \frac{1}{c} W \psi_y + i \, p_z \psi_x - i \, p_x
\psi_z = 0,
\\ \\
\displaystyle \frac{1}{c} W \psi_z + i \, p_x \psi_y - i \, p_y
\psi_x = 0, \ea \right. \qquad \qquad \underline{p_x \psi_x + p_y
\psi_y + p_z \psi_z = 0,}
\]

${}$

\noindent [and the Maxwell equations may be written in the compact
form, analogous to that of the Dirac equation,]

\[
\left[ \frac{1}{c} \, W + \left( \alpha , p \right) \right] \psi =
0,
\]

${}$

\noindent [with]

\[ \alpha_x \; = \; \left( \ba{ccc}
0 & 0 & 0 \\
0 & 0 & -i \\
0 & i & 0 \ea \right) , \qquad \alpha_y \; = \; \left( \ba{ccc}
0 & 0 & i \\
0 & 0 & 0 \\
-i & 0 & 0 \ea \right) , \qquad \alpha_z \; = \; \left( \ba{ccc}
0 & -i & 0 \\
i & 0 & 0 \\
0 & 0 & 0 \ea \right) .
\]

${}$

\noindent [In order that the equation satisfied by $\psi$ have a
non trivial solution, it needs that the energy $W$ is given by]

\[
\frac{W}{c} = \left( \ba{ccc}
0 & i p_z & -i p_y \\
-i p_z & 0 & i p_x \\
i p_y & -i p_x & 0 \ea \right) ,
\]

${}$

\noindent [whose eigenvalues are]

\[
\frac{W}{c} =  p , \qquad \frac{W}{c} = - p , \qquad \frac{W}{c} =
0 ,
\]

${}$

\noindent [while the ratio between the components of the
eigenvector corresponding to the null eigenvalue is given by:]

\[
\psi_x : \psi_y : \psi_z \; = \; - p_x^2 : -p_x p_y : -p_x p_z \;
= \; p_x : p_y : p_z,
\]

\[
\widetilde{\psi} \psi = E^2 + B^2 .
\]

\bigskip \bigskip

\centerline{\sc {\it Excerpta} Senatore IIc}

\bigskip

\noindent Oppenheimer has tried to build such equations with an
inductive procedure by using the experimental properties of the
photons. \\
These properties may be summarized in the following way: \\
1) the photons move at the speed of light; \\
2) the energy and the momentum are related each other by the very
simple relation: $W = c p$; \\
3) given the strength and the direction of the momentum, two
possible polarization states are present; \\
4) the photon has a spin that may assume the values $\displaystyle
\pm \frac{h}{2 \pi}$ along the propagation axis. For an
experimental justification of this last postulate, we observe
that, when an atom emits some radiation, its angular momentum
around a given $z$ axis changes from $+$ to $\displaystyle -
\frac{h}{2 \pi}$ (or $0$) depending on the fact that one or the
other of the following transitions takes place:
\[
m \rightarrow m^\prime = m \mp 1
\]
\[
\left( m \rightarrow m^\prime = m \right) .
\]
Then, if we observe a photon along the $z$ axis, we can certainly
say that such a photon cannot be due to the last transition, since
the intensity of this vanishes along the [$z$] axis. On the other
hand if the photon lies exactly on the $z$ axis, this cannot
depend on the angular momentum around this axis, so that from the
conservation of the angular momentum of the system we must admit
the existence of the spin of the photon, in a way that it may
assume the two values $\displaystyle \pm \frac{h}{2 \pi}$ along
the propagation direction. Such an interpretation of the spin of
the photon has been proposed for the first time by Dirac, when he
was not yet R.R.S.\footnote{Probably such initials should be
corrected in F.R.S., {\it Fellow of the Royal Society}. In this
way, the author points out that the work by Dirac was performed
before his election as a member of the Royal Society of London (in
March 1930). In any case, it is quite interesting to find such an
historical note in a scientific manuscript by Majorana.} \\
Since we are concerned with two polarized components, it would be
worth to consider a two-component wave theory. Indeed we can build
some equations of the desired form satisfying to the first three
postulates. These equations are the following ones:
\[
\left[ \frac{W}{c} + \left( \sigma , p \right) \right] \psi = 0 ;
\qquad \qquad \sigma_x = \left| \ba{cc} 0 & 1 \\ 1 & 0 \ea \right|
, \quad \sigma_y = \left| \ba{cc} 0 & -i \\ i & 0 \ea \right| ,
\quad \sigma_z = \left| \ba{cc} 1 & 0 \\ 0 & - 1 \ea \right| .
\]
I point out that these equations are nothing other that half of
the Dirac equation.




\begin{thebibliography}{99}

\bibitem{Recami}
E. Recami, {\it Il caso Majorana} (Oscar Mondadori, Milan, 1991,
second edition); {\it idem} (Di Renzo, Rome, 2002, fourth revised
and enlarged edition).

\bibitem{pw}
S. Esposito, Physics World {\bf 19} (2006) 34,

\bibitem{DeGregorio}
A. De Gregorio and S. Esposito, 
Sapere, Giugno 2006, p. 56 [arXiv:physics/0602008]; A. De Gregorio
and S. Esposito, 
[arXiv:physics/0602146].

\bibitem{Moreno}
S. Esposito, 
Nuovo Saggiatore {\bf 21} (2005) 21; 
A. Drago and S. Esposito, Phys. Persp., in press
[arXiv:physics/0503084].

\bibitem{Weyl}
A. Drago and S. Esposito, 
Found. Phys. {\bf 34} (2004) 871.

\bibitem{Senatore}
G. Senatore, Lecture, Department of Physical Sciences, University
of Naples ``Federico II'', March 30, 1998.

\bibitem{bibliopolis87}
E. Majorana, {\it Lezioni all'Universit\`a di Napoli},
(Bibliopolis, Naples, 1987).

\bibitem{bibliopolis06}
E. Majorana, {\it Lezioni di Fisica Teorica}, edited by S.
Esposito (Bibliopolis, Naples, 2006).

\bibitem{path}
S. Esposito, Eur. J. Phys. {\bf 27} (2006) 1147
[arXiv:physics/0603140]; S.
Esposito, 
[arXiv:physics/0512174].

\bibitem{Hei29}
W. Heisenberg, Naturwissenschaften {\bf 26} (1929) 490. 

\bibitem{Jordan26}
P. Jordan, {\it Anscauliche Quantentheorie eine Einf\"uhrung in
die moderne Auffassung der Quantenenerscheinurgen} (Springer,
Berlin, 1926).

\bibitem{Dirac27}
P.A.M. Dirac, Proc. Roy. Soc. London {\bf 114} (1927) 243, 
710. 

\bibitem{JK27}
P. Jordan and O. Klein, Z. Phys. {\bf 45} (1927) 755. 

\bibitem{JW28}
P. Jordan and E. Wigner, Z. Phys. {\bf 47} (1928) 631. 

\bibitem{HP29}
W. Heisenberg and W. Pauli, Z. Phys. {\bf 56} (1929) 1, 
{\it ibidem} {\bf 59} (1930) 168. 

\bibitem{Waller30}
I. Waller, Z. Phys. {\bf 61} (1930) 837. 

\bibitem{Dirac30}
P.A.M. Dirac, Proc. Roy. Soc. London {\bf 126} (1930) 360; 
Proc. Cambridge. Phil. Soc.  {\bf 26} (1930) 361. 

\bibitem{Weyl29}
H. Weyl, Z. Phys. {\bf 56} (1929) 330. 

\bibitem{Pauli33}
W. Pauli, Handbuch der Physik, {\bf 24} (1933) 83. 
See also the letters by W. Pauli to P.M. Blackett (19 April 1933)
and to P.A.M. Dirac (1 May 1933) reported in W. Pauli, {\it
Wissenschaften Briefwechsel mit Bohr, Einstein, Heisenberg, U.A.
II: 1930-1939} (Springer, Berlin, 1985).

\bibitem{HS33}
See the letter by W. Heisenberg to A. Sommerfeld (17 June 1933)
reported in W. Pauli, {\it Wissenschaften \dots}, {\it loc. cit.}

\bibitem{elpos}
E. Majorana, 
Nuovo Cim. {\bf 14} (1937) 171. 

\bibitem{Hei31}
W. Heisenberg, Ann. der Physik {\bf 10} (1931) 888. 

\bibitem{Quaderni}
See the list of the scientific unpublished manuscript reported in
E. Recami, 
Quaderni di Storia della Fisica, {\bf 5} (1999) 19.

\bibitem{Opp31}
J.R. Oppenheimer, Phys. Rev. {\bf 38} (1931) 725.

\bibitem{RecEspGian}
R. Mignani, E. Recami and M. Baldo, Lett. Nuovo Cim. {\bf 11}
(1974) 568; E. Giannetto, Lett. Nuovo Cim. {\bf 44} (1985) 140 and
145; E. Giannetto, in F. Bevilacqua (ed.) {\it Atti IX Congresso
Naz.le di Storia della Fisica} (Milan, 1988) p.173. S. Esposito,
Found. Phys. {\bf 28} (1998) 231. 

\bibitem{atomi}
E. Majorana, Nuovo Cim. {\bf 9} (1932) 43. 

\bibitem{sdoppiamento}
G. Gentile and E. Majorana, Rend. Acc. Lincei {\bf 8} (1928) 229.

\bibitem{RossiFermi}
E. Fermi and B. Rossi, Rend. Acc. Lincei {\bf 17} (1933) 346.

\bibitem{FNM}
E. Fermi, {\em Collected Papers}, edited by E. Amaldi {\it et
al.}, University of Chicago Press - Accademia Nazionale dei
Lincei, Chicago - Rome, 1962-1965, 2 volumes, Vol. I (1962).

\bibitem{allunion}
M.P. Bronshtein, V.M. Dukel'skii, D.D. Ivanenko and Yu. B.
Khariton (eds.), {\it Problems of Modern Physics Vol. 24: The
Atomic Nucleus: Collected Papers of the First All Union Nuclear
Conference, held 24-30 September 1933 at Leningrad}, Leningrad and
Moscow State Technical-Theoretical Publishing House, 1934.

\bibitem{Vizgin}
V.P. Vizgin, Physics Uspekhi {\bf 42} (1999) 1259. 

\bibitem{positrona}
P.A.M. Dirac, in M.P. Bronshtein, V.M. Dukel'skii, D.D. Ivanenko
and Yu. B. Khariton (eds.), {\it loc. cit.}, p. 129. 

\bibitem{Segre}
E. Segr\'e, {\em Enrico Fermi Physicist} (The University of
Chicago Press, Chicago, 1970).

\bibitem{FNM77b}
See the date reported in E. Fermi and G.E. Uhlenbeck,
Phys. Rev. {\bf 44} (1933) 510. 

\bibitem{kern}
E. Majorana, Z. Phys. {\bf 82} (1933) 137; 
Ricerca Scientifica {\bf 4} (1933) 559. 

\bibitem{degneutro}
A. De Gregorio, Stud. Hist. Philos. Mod. Phys. {\bf 37} (2006)
330; 
A. De Gregorio, [arXiv:physics/0603261].

\bibitem{Tamm}
E. L. Feinberg, Physics Uspekhi {\bf 38} (1995) 773. 

\bibitem{cold}
G. Waysand, Phys. Stat. Sol. {\bf C2} (2005) 1566. 

\bibitem{Weisskopf}
V.F. Weisskopf, talk given at the Erice Summer School in
High-Energy Physics, Erice (Italy), 1971.

\bibitem{Peierls}
See the Chronology of the Life of Sir Rudolph Ernst Peierls in
R.H. Dalitz and R. Peierls, {\it Selected Scientific Papers of Sir
Rudolf Peierls with commentary} (World Scientific - Imperial
College Press, Singapore - London, 1997).

\end{thebibliography}
\end{document}